\input aa.cmm
%
%
\newcount\fignumber\fignumber=1
\def\nfig{\global\advance\fignumber by 1}
\def\fignam#1{\xdef#1{\the\fignumber}}
\fignam{\FA} \nfig 
\fignam{\FB} \nfig 
\fignam{\FC} \nfig 

\voffset=-1 truecm
\overfullrule=0 pt

\MAINTITLE{The role of references in the astronomical discourse} 

\AUTHOR{R. Girard and E. Davoust}

\INSTITUTE{
UMR 5572, Observatoire Midi-Pyr\'en\'ees, 
14 Avenue E. Belin, 31400 Toulouse, France,
}

\DATE{Received September 3, accepted December 16, 1996}

\ABSTRACT{
We have counted the number of references in 1179 papers published in {\it 
Astronomy and Astrophysics} over twenty years.  The number of references has 
increased by 60\% between 1975 and 1995, reflecting the increase (by the same 
amount) of the literature which must be cited, and of the number of pages per 
paper. There are 1.5 times more references in predominantly observational 
fields than in others. References are used 1.65 times in the text, and there 
is no trend with time or field.  They appear mostly in the introduction (30\%) 
and in the main body of the paper (60\%), but papers in predominantly 
observational fields tend to use less references in the introduction and more 
in the main body than papers in the other fields.  Most references (62\%) 
serve to support a result, and tend to be of theoretical nature.  Astronomers 
are a very conformist bunch, as there are no trends with nationality, and 
references to conflicting evidence are kept at about 8\%. The analysis of a 
series of papers by de Vaucouleurs on the Hubble constant shows how a 
controversial subject affects the use of references.} 


\THESAURUS{ } %

\maketitle %

\titlea{Introduction} 

References have often been used by sociologists of science for measuring the 
output, impact and quality of scientific research.  The number and frequency 
of references in astronomical papers have been extensively studied, for 
example for evaluating the impact of research by various groups of astronomers 
(Jaschek 1992, Trimble 1985, 1993ab, Peterson 1987), the productivity of 
telescopes (Abt 1980, 1985), the life-time of papers (Abt 1981, Peterson 
1988), the effect of funding on the quality of research (Abt 1984), the flows 
of information between astronomy and neighboring fields (Davoust et al. 1993). 

But little attention has been paid, at least in astronomy, to the way 
references are used in the scientific discourse inside the paper.  They play 
an essential role in supporting the authors' claim that the presented research 
is correct and valuable.  Thus exposing the method by which astronomers try to 
reach this goal is very useful, as it provides a manual of referencing style 
that one should follow in order to be published. 

In this paper, after presenting our sample (Sect. 2), we analyze how and why 
the number of references varies with time and scientific field (Sect. 3, 4), 
we then study the distribution of references in the body of the papers (Sect. 
5), determine the different purposes for which they are used in the text 
(Sect. 6), and analyze how a controversial subject affects their use (Sect. 
7).  The conclusion provides a manual of referencing style for astronomers. 

\titlea{The sample of papers and references}

We chose {\it Astronomy and Astrophysics} (hereafter A\&A) as the source of 
references. This journal is the medium in which most European astronomers 
publish, and one of the most distinguished journals in its domain.  It covers 
most fields of astronomy, and has the advantage that the papers are sorted 
into 8 or 9 fields in its table of contents.

Although A\&A started publication in 1969, we decided to count references from 
1975, considering that, by that date, the journal had found its cruising 
speed.  We sampled references every five years, a small enough interval to 
determine the time dependence of the reference patterns.
Four volumes (400 to 600 pages each) of A\&A per year was enough to get good 
statistics, with small fluctuations (see Table 1). In 1995, we only used three 
volumes (which contain 800 pages each).  

The number of fields increased over the years, reflecting the expansion of 
research in most subjects. In 1975, fields 7 and 8 were only one field.  In 
1980, field 5 was broken into two subfields (stellar structure and evolution; 
stellar atmospheres), and in 1984, fields 6 and 7 were both divided into two 
subfields.  To save ourselves some work, we sorted by hand the papers of 
fields 7 and 8 in 1975, but did not consider the two subfields of 5, 6 and 7 
as separate.  This is justified because of the close links between the 
subfields. The field numbers are identified in Table 1. 

We counted by hand all the references at the end of all papers of the 
first four volumes of 1975, 1980, 1985, 1990, and of the first three volumes 
of 1995 of A\&A. We excluded letters and research notes from the counts, 
because these shorter papers have a different purpose than more complete 
papers, and thus presumably a different style and rhetorical method.  We 
also excluded the {\it Supplement Series} of the journal, for similar reasons. 
 
\titlea{Evolution of the number of references per paper}

The number of references per paper has increased in all fields over the 
years.  The rate of increase is of the order of 60\% over 20 years.  The fact 
that the world literature has also increased by 60\% over that period of time 
(estimated by the number of entries in the semi-annual issues of {\it 
Astronomy and Astrophysics Abstracts}) provides an obvious explanation for 
this trend : there are simply more papers to cite. 

The number of pages (normalized to 1000-word pages) per paper has also 
increased at almost the same rate (55\%), from 6.6 to 10.2 pages.  The 
tendency for the number of pages per paper to increase is in fact a general 
one; Trimble (1984) gives the following figures for the rate of page increase 
between 1950 and 1980-83 : mathematics (77\%), physics (27\%), chemistry 
(93\%), and astronomy (82\%).  Returning to our subject, the net result is 
that the number of references per normalized page in A\&A has remained fairly 
constant at 3.4$\pm$ 0.2.  In other words, the absolute number of references, 
not their density, has increased over the years.  There are more published 
results to cite and discuss when writing a paper, if one wants to be 
comprehensive, and this requires more space in the journal. The interesting 
implication here is that the number of {\it relevant} references has increased 
in proportion with the total number of references, and it is somewhat 
reassuring to find an indication that the inflation in papers does not result 
from an excess of useless ones. 

Abt (1987) has found a tight correlation between the number of references {\it 
nr} and paper length {\it pl} (normalized to 1000-word pages) of astronomical 
papers published in 1986, such that $ nr = 9.9 + 2.18 pl$. This relation 
predicts the correct number (slightly overestimated, but within one unit) of 
references per paper in A\&A for all years except for 1995, where it is 5 
references short. 

These trends are shown on Fig. 1, where minor differences in slope of the 
three curves may be explained by editorial constraints and statistical 
fluctuations. The number of references counted by Vin (1995) in 6 astronomical 
journals (A\&A, A\&AS, ApJ, ApJS, MNRAS and AJ) in 1980-93 and by Abt (1987) 
in 8 astronomical journals in 1986 are also plotted (as crosses), for 
comparison.  They fall very close to the solid line, which means that our 
sampling of A\&A is representative of the world astronomical literature. 

\begfig 5cm
\figure{\FA}
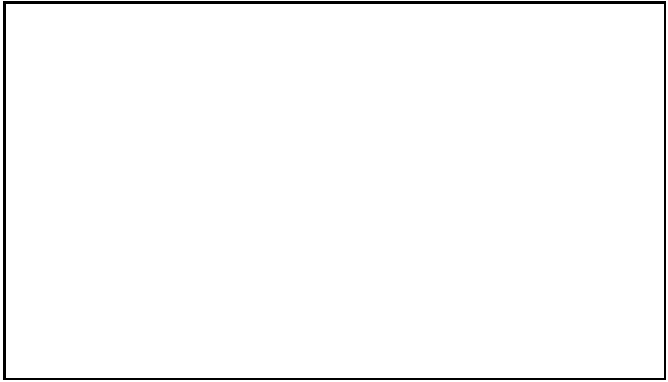
{Evolution of the number of references per paper in A\&A (dotted line), of 
papers per year in {\it Astronomy and Astrophysics Abstracts} (in thousands, 
solid line) and of normalized pages per paper in A\&A (dashed line). The 
number of references per paper in 6 or 8 astronomy journals is also given for 
a few years (crosses)} \endfig 

Field 9 (physical \& chemical processes) stands out in this respect, because 
the number of references has been decreasing since 1985 (see Fig. 2). We 
checked that this is not a statistical fluctuation, as the tendency is even 
more marked when including more papers per year.  This may be due to the fact 
that this is a borderline field of astronomy; astronomers perhaps cannot keep 
up with the information explosion in neighboring fields. 

\titlea{Patterns of reference in the different fields}

While the number of references per paper increases with time, it is not 
uniform over all fields in a given year. There are two categories of fields, 
those with a high rate of citation (stars, galactic and extragalactic 
astronomy, interstellar matter), with an average of 42 references/paper in 
1995, and those with a low rate of citation (the Sun, celestial mechanics, 
instruments, planetary systems, physical \& chemical processes), with an 
average of 27 references/paper in 1995.  Figure 2 shows the number of 
references per paper in the different fields at the five epochs, and 
the actual numbers are given in Table 1, where the last line is the mean 
weighted by the number of papers in each field. 

\begfig 5cm
\figure{\FB}
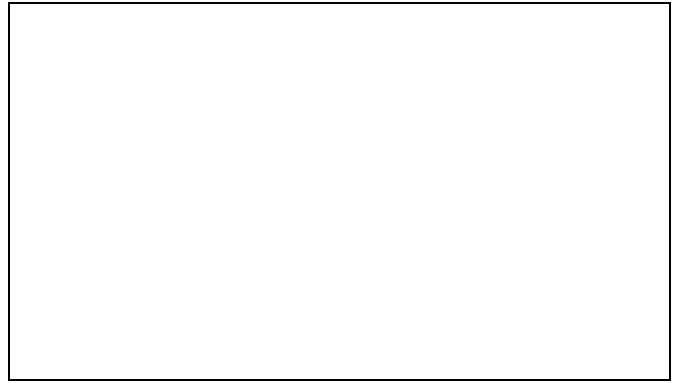
{Distribution of the number of references per paper in each field in 1975 
(dotted line), 1980 (dashed line), 1985 (longdashed line), 1990 (dot-dashed 
line), 1995 (dot-longdashed line), and of the proportion of observational 
papers over all epochs (in percent, solid line).  The field numbers are 
identified in Table 1} 
\endfig 

We tested two possible assumptions for explaining this trend.

\begtabfullwid
\tabcap{1}{Number of references per paper and per field in A\&A 
over 20 years. The last column is the total number of papers}
\halign{#\hfill&\quad#\hfill&\quad\hfill#&\quad\hfill#&\quad\hfill#
&\quad\hfill#&\quad\hfill#&\quad\hfill#\cr
\noalign{\medskip\hrule\smallskip\hrule\medskip}
 & Subject & 1975& 1980& 1985& 1990& 1995&n\cr
\noalign{\medskip\hrule\medskip}
1& The Sun& 22.16& 20.88& 24.21& 26.68& 30.04&98\cr
2& Planetary Systems& 18.83& 25.70& 28.27& 28.47& 33.29&66\cr
3& Celestial Mechanics \& Astrometry& 10.83& 11.33& 16.67& 19.00& 26.38&36\cr
4& Instruments \& Data Processing& 8.57&  7.70& 15.18& 14.40& 21.50&51\cr
5& Stars& 26.81& 25.91& 28.53& 33.22& 38.75&352\cr
6& Cosmology \& Extragalactic Astronomy& 25.94& 27.02& 35.73& 34.80& 43.96&227\cr
7& Galactic Structure \& Dynamics& 30.07& 29.65& 36.96& 29.00& 43.88&106\cr
8& Interstellar Medium& 29.32& 34.17& 36.08& 34.04& 39.90&143\cr
9& Physical \& Chemical Processes& 16.50& 18.75& 25.69& 23.00& 22.20&100\cr
\noalign{\medskip\hrule\medskip}
 & Mean & 24.07 & 24.65 & 30.15 & 30.45 & 37.20\cr                                                                                                   
\noalign{\medskip\hrule\smallskip\hrule\medskip}
}
\endtab

We first checked whether the difference could be due to the internal dynamics 
of the fields, if there are more references per paper in the fields which 
contain more papers.  The numbers of papers published in the different fields 
have been counted by Davoust \& Schmadel (1987, their Table VI) in the period  
1969-1985.  Using their data, we found that there is no correlation between 
the number of published papers and the number of cited references per paper in 
the different fields.  Take for example fields 1 to 4, which have a low rate 
of citations; if our assumption is correct, there should be the same (low) 
number of published papers in these four fields.  But there are at least twice 
as many papers over 27 years in field 2 than in 1 and in 4 than in 3.  
Conversely, there should be the same (high) number of published papers in 
fields 5 and 8; but there are in fact over twice as many in 5 than in 8.  
This trend is also apparent in the last column of our Table 1, although with 
larger fluctuations, since we have been taking small samples. 

This lack of correlation suggests that papers do not necessarily cite 
references in their own field, provided that the proportion of meaningful 
papers (the ones that deserve to be cited) is the same in all fields. 
An interesting implication, which still has to be checked, is that 
some fields, 8 for example, must be strongly related to other fields
(supplying information to, or drawing it from, other fields), since 
they apparently cite more papers than there are papers to cite in their own 
field. They are thus more multidisciplinary than others, which in our opinion 
is a quality. 

We then checked a second possibility, which turned out to be the correct one, 
namely that the difference in citation rate was due to the type of research, 
theoretical or observational, predominant in the different fields. We sorted 
all the papers of our sample into one of three categories, theory, 
observations, or theory-and-observations, and redistributed the numbers of the 
last category evenly between the two others.  

Theoretical papers outnumber observational ones by 2 to 1, and the 
ratio remains constant (0.64$\pm$0.02) over 20 years.  But there is a definite 
trend with field (see Fig. 2), as there are less references per paper in the 3 
fields with a very strong majority of theoretical papers (celestial mechanics, 
instruments \& data processing, physical \& chemical processes). A possible 
explanation is that theoretical papers only cite theoretical ones, whereas 
observational ones, which must provide interpretation or modeling to be in the 
main journal, cite both types of papers.  Another explanation is that 
observational papers benefit from several large data bases which provide 
immediate access to most references related to a given celestial object
(we are indebted to Carlos Jaschek for providing this remark).

The correlation between field and rate of citation is not as sharp as the 
one between field and type (theoretical vs observational), because in all 
papers, irrespective of the field, the references are mostly to theoretical 
papers (rather than to observations, methodology or for other purposes; see 
Sect. 6).  
                                           
The number of references per paper has also been studied in other fields 
of science. Vin (1995) gives the following numbers for the year 1993 : 
computational sciences (22.2), mathematics (22.5), physics (26.3), chemistry 
(39.5), biochemistry and molecular biology (44.1).  But these numbers are not 
directly comparable to those in astronomy, because the average length of 
papers may not be the same in the different fields; Abt (1987) notices that, 
in 1986, astronomy papers were 1.8 times longer than physics papers. 

We found no trends related to the nationality of the authors. Abt (1987) 
reaches a similar result; he notes that there is no difference between 
American astronomical journals and others (namely MNRAS, A\&A and PASJ) in 
their average number of references per paper of same length.  His conclusion 
is that the scientific methods are the same and that authors tend to copy the 
styles of the papers that they read.  We may add that the norm that 
astronomers have to conform to in order to have their papers accepted for 
publication is rather strict and enforced by referees from all over the world.  
The fact that the overwhelming majority of papers are in English may also 
contribute to this uniformity.  

\titlea{Distribution of the references in the text}

After identifying variations in the number of references per paper with time 
and field, we now analyze how the references are used in the paper itself.

References are used on the average 1.65$\pm$0.13 times in the text, and there 
is no trend with time or field.  In other words, the typical reference is not 
used repeatedly in the discourse, but only once or twice.  The case where up 
to 3 or 4 references are frequently used in the text does occur, but 
relatively rarely (20 to 30\% of the time).  In other words, authors do not 
tend to overemphasize the relations of their paper to the literature. 

The references are used in increasing proportion in the abstract (1.1\%), the 
conclusion (8.2\%), the introduction (30.2\%), and the main body of the paper 
(60.5\%). These rates reflect to some extent the proportion of text in the 
different sections, except for the introduction, which is obviously an 
important part of the paper, where the scientific context is traditionally 
presented, with an appropriate number of references.  

There are slight differences with fields, as there are relatively more 
references (9 points in \%) in the main body (and conversely relatively fewer 
references in the introduction) of papers in predominantly observational field 
than in that of papers belonging to more theoretical fields.  This simply 
means that the excess of references which characterizes predominantly 
observational fields (see Sect. 4) is cited in the course of the discussion, 
rather than in the introduction.  Since this excess is made up of 
references to observations (see Sect. 6), this also implies that observations 
are cited in the discussion rather than in the introduction.

Celestial mechanics stands out in this respect, as there are three times more 
references in its abstracts than in those of other fields. As the abstract is 
the most visible part of a paper, this probably reflects the strong need of 
authors in celestial mechanics to emphasize the relevance of their work. 
Another striking characteristic of this field is that the references are often 
to unusually old papers. These should perhaps be considered specific features 
of academic subjects. 

\titlea{The purpose of references in the astronomical discourse}

Our main interest is to investigate for what purpose references are cited in a 
given paper. To this end, we adopted a simple classification of the references 
into 6 purposes, which are listed in Table 2. The distinction between purpose 
2 and 3 is sometimes subtle; similar research is generally cited in the 
introduction, without any judgment on its contents. We further subdivided 
purpose 3 and 4 (after merging them together) into reference to theoretical, 
observational or methodological information.  The distinction between 
theoretical and methodological reference is not always obvious; the latter is 
in principle a reference to theoretical means (an equation, a model) for 
reaching one's goal.  The references to observational means (instruments, 
data) to be used in the discourse are classified as purpose 6.

\begtabfullwid
\tabcap{2}{Frequency of references (in percent) according to purpose in the 
text, in the 9 fields of astronomy.  The last three lines are the 3 types of 
(supporting and conflicting) information}   
\halign{#\hfill&\quad#\hfill
&\quad\hfill#&\quad\hfill#&\quad\hfill#&\quad\hfill#&\quad\hfill#
&\quad\hfill#&\quad\hfill#&\quad\hfill#&\quad\hfill#&\quad\hfill#\cr
\noalign{\medskip\hrule\smallskip\hrule\medskip}
&Purpose $\backslash$ Field &1&2&3&4&5&6&7&8&9&mean\cr
\noalign{\medskip\hrule\medskip}
1 &Scientific, historical context&9.16&10.61&19.24&13.10&10.07&9.49&5.20&2.91&13.84&10.4\cr 
2 &Similar research&11.06&7.52&19.70&19.05&13.49&15.51&13.69&2.68&6.59&12.1\cr 
3 &Supporting information&67.57&76.55&40.79&52.38&57.93&57.25&65.53&72.48&67.99&62.1\cr 
4 &Conflicting information&7.83&2.56&5.31&0.00&14.47&10.73&5.67&21.47&8.61&8.5\cr 
5 &Definition&1.79&1.80&2.08&4.76&1.52&0.00&3.03&0.46&0.00&1.7\cr 
6 &Source of data, instruments &2.60&0.95&12.88&10.71&2.53&7.02&6.87&0.00&2.97&5.2\cr 
A &Theory&51.31&62.34&51.59&75.00&46.94&47.19&46.07&45.18&52.56&53.1\cr 
B &Observations&32.20&27.19&22.07&20.45&33.85&42.87&39.77&50.31&32.69&33.5\cr 
C &Methodology &16.49&10.46&26.34&4.55&19.20&9.93&14.16&4.51&14.75&13.4\cr 
\noalign{\medskip\hrule\smallskip\hrule\medskip}
}
\endtab

Taking the present paper as an example, we would classify the references as 
follows.  In the introduction, all references are 1. In Sect. 3, the 
first occurrence of Abt (1987) is 3C, because we did not parametrize our 
correlation, otherwise it would be 3A; Trimble (1984), Vin (1995) and the 
second occurrence of Abt (1987) are 3B. In Sect. 4, Davoust \& Schmadel (1987) 
is 6; Vin (1995) is 2, since we cannot compare the quoted numbers to ours, 
otherwise it would be 3B; the first occurrence of Abt (1987) is 6, since we 
implicitly use the quoted number of pages to argue that no comparison is 
possible in general, otherwise it would be 3B; the second occurrence of Abt 
(1987) is 3B. In Sect. 7, all references are 6. 

\medskip

We thus classified all the references in the papers of 1995 in our sample. The 
results are displayed in Table 2 and Fig. 3, and discussed next. 

--- Most references (62\% on the average, or 65\% excluding celestial 
mechanics) are to information in support of the results of the paper.  This is 
not an unexpected result, our only merit is to quantify the effect.  Celestial 
mechanics stands out with a low 40\%, because the context is twice as 
important as in other fields. 

--- All the other types of references are around or below 10\%.

--- References to conflicting information fluctuate around 8\%.  There is of 
course a bias here, since a piece of research with a majority of conflicting 
evidence would be very difficult to publish.  We believe that a large number 
of citations to conflicting evidence is a sign of good health, and that the 
lack of controversies is evidence for a static field, with no major 
breakthrough in progress; in this respect, fields 5 and 8 rate very 
positively. A remarkable fact is that field 8 (the interstellar medium) has 
very high rates of supporting {\it and} conflicting references, but, 
since there are very few references for other purposes, the absolute numbers 
are not unusually large.  Field 8 still strikes us as an unusual one, whose 
referencing pattern should perhaps be investigated further. 

--- References to context and to similar research only amount to 22\%.  
Because they are usually confined to the introduction, we expected their 
relative number to be of the order of 30\%, which is the proportion of 
references cited in the introduction (see Sect. 5). The reason for this 
discrepancy is that the scientific debate often already starts before the end 
of the introduction, with references (generally) in support of the project. 

\medskip

We now turn to the relative rates of theoretical, observational and 
methodological references.

--- The proportion of references to theoretical papers is fairly constant, with 
an average value of 53\%.  Fields 2 (62\%) and 4 (75\%) are the most deviant 
ones. 

--- The proportion of references to observational papers depends on the field 
in roughly the same way as the absolute number of references (see Fig. 1), 
which in turn depends on the type (observational or theoretical) of the field. 
If we take for example fields 5 to 8, which are predominantly observational, 
they produce 1.56 times more references than the other fields (see Sect. 4), 
and these references are 1.54 times more observational than in the other 
fields. In other words, there is a good correlation between the types of the 
citing and cited papers. 

\begfig 5cm
\figure{\FC}
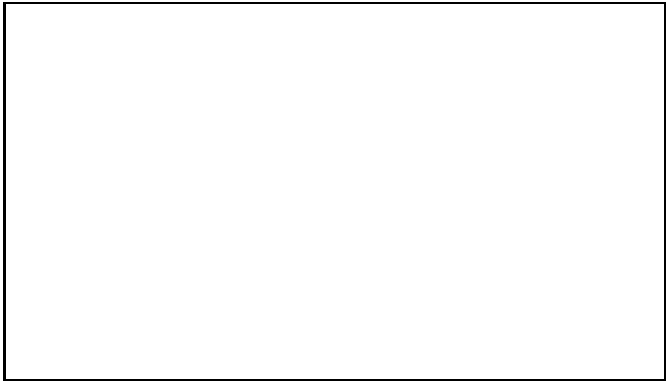
{Relative importance of the different purposes of references in the 9 fields 
of astronomy.  Context (longdashed line), similar research (dotted line), 
supporting information (dashed line), conflicting information (solid line),  
definition (dot-dashed line), source of data or instruments (dot-longdashed 
line)
}
\endfig 

\titlea{A case study : de Vaucouleurs on the Hubble constant}

G\'erard de Vaucouleurs (1918-1995) is one of the most remarkable astronomers 
of the second half of this century; he was highly productive, with 208 papers 
published between 1969 and 1985 (Davoust \& Schmadel 1987), and highly cited, 
with 686 citations in 1982 (ranking second only to Chandrasekhar; Trimble, 
private communication; Trimble 1985).  Among other noteworthy achievements, he 
played a fundamental role in the debate on the Hubble constant.  These enviable 
qualities led us to select his series of 8 papers on the Hubble constant which 
appeared in the {\it Astrophysical Journal} between 1977 and 1993 (see de 
Vaucouleurs 1993, de Vaucouleurs \& Bollinger 1979 and references therein) as 
a basis for a case study on rhetorical skill.  Readers may object that the 
journal is not A\&A, but this series of papers is simply unique.  Who else 
would have written a homogeneous set of 8 papers or more in the past twenty 
years on one of the major contemporary issues in astronomy? 

The total number of references in this series of papers is slightly above 
average (50$\pm$15 per paper), and their frequency of appearance in the text 
is well above average.  Paper VIII, which appeared 14 years after the first 
seven, and is de Vaucouleurs' final word on the value of the Hubble constant, 
contains 72 different references, but they appear 466 times in the text (a 
frequency of 6.5 per reference).  Excluding this final paper, the frequency is 
2.0 per reference (compared to 1.65 on the average in A\&A). 

Keeping in mind that these papers are essentially a debate between de 
Vaucouleurs' determination of the Hubble constant and that of Sandage \& 
Tamman (hereafter ST), we find that autocitations and references to ST 
appear with the same high frequency in the text (2.5, excluding paper VIII). 
In this last compilation paper, references to other papers have a higher 
frequency (8.7) than autocitations (5.2) and references to ST (4.0).  
Excluding autocitations, references to ST, and paper VIII, we find that the 
frequency of references in the text is normal (1.7).  It is thus the debate 
with ST which raises that frequency. It is interesting to note that, in the 
reference lists, the number of autocitations rises monotonically from 5 in 
paper I to 32 in paper VII, while the number of references to ST remains quite 
constant, between 5 and 7. The author thus increasingly drums his other papers 
to the fore as the debate progresses.   

The division of the discourse into 8 papers is the main cause for the unusual 
distribution of references in the text.  The introduction is always very short 
and contains a very small proportion of references (6\%), as most of them are 
used to structure the debate in the text (72\%) and in the appendices 
(17.2\%).  Most papers have appendices, this is part of de Vaucouleurs's 
style which often relies on them as well as on footnotes; the discourse thus 
gains in conciseness and clarity or, conversely, the paper thus gains in 
precision and completeness.  But in the present series, the number of 
appendices and footnotes (respectively 2 and 6 per paper on the average) is 
unusually high. 

The intense (but courteous) debate modifies the proportions of references for 
different purposes.  Only 40\% of the references are to supporting information 
(including autocitations) compared to 62.1\% on average, because conflicting 
information (15\% instead of 8.5) and sources of data (25\% instead of 5.2\%) 
are more cited than usual.  The author is obviously in ennemy territory, and 
draws more heavily on data to make up for friendly forces.  

Data can be cited, but they can also be displayed in figures and tables. The 
number of figures increases progressively in the series, indicating 
more frequent recourse to visual display of information as the discussion 
proceeds, a wise strategy since figures are generally easier to perceive than 
numbers. There are also more tables in the second half of the series than in 
the first half, but no striking progression. 
  
Finally, leaving the approach of scientometrics, the reader cannot but be awed 
at the talent with which the author leads him through the intricacies of the 
scientific debate, keeping him in suspense by only revealing the adopted value 
of the Hubble constant in paper VII.  

\titlea{Conclusion}

The statistical analysis of the number and distribution of references in the 
text of 1179 papers sampled from A\&A over 20 years provides a manual of 
referencing style for astronomers. 

The number of references per paper has steadily increased over the years, in 
pace with the increase in the number of papers published worldwide, and with 
the consequent increase in the number of pages per paper.  One striking 
exception is the number of references in the field of physical and chemical 
processes, which has been decreasing since 1985.  Astronomers are expected to 
keep up with the inflation in astronomical literature at the expense of that 
in neigboring fields, and to cite it accordingly.  

But the number of references per 1000-word page remains constant (3.4$\pm$0.2 
references/page), as well as the number of times a reference is cited in a 
paper (1.65 times on the average).  The number of references may increase 
with time, but they should be cited with moderation.  No more than 60\% of the 
references should be used to support the claims of the paper, and, for the 
sake of credibility, about 10\% of the references should provide conflicting 
evidence. 

The number of references per paper depends on the type (predominantly 
theoretical or observational) of the field.  Authors may use 1.5 times more 
references if their paper belongs to the latter type of field.   

The introduction of the paper plays a special role, with a large number of 
references (30.2\% in relative numbers) compared to its size.  One third of 
them should provide information for (and, once in ten times, against) the 
specific project of the paper. 

There is no trend in referencing pattern with the nationality of the 
authors.  Conformity is the rule.

The analysis of de Vaucouleurs's series of papers on the Hubble constant shows 
that, in papers at research fronts, more space than average should be given to 
data and conflicting arguments.  Authors may then appeal to more data than 
usual and raise their autocitation rate in the text to defend their point of 
view. 

The above advice should of course not be taken seriously as such. It is 
simply a summary of our results.

Finally, we give a few suggestions for possible follow-up studies.
We did not study systematically the age distribution of the cited 
references, and how it varies with time in the different fields
(we are thankful to the referee for pointing this out to us). Such 
distributions deserve to be determined, as they are indicative of the long- 
and short-term dynamics of the various fields.  Another interesting study to 
be undertaken is to understand why some fields, like the interstellar medium, 
seem to generate more references to other fields than to its own, whether this 
is an artefact of the classification system, or a true effect. 

\acknow{G\'erard de Vaucouleurs certainly did not expect that his papers would 
be used in this way when he handed a set of reprints to one of us many 
years ago, nor would he have approved.  But this study was done with due 
respect to his memory, and in praise of his research and teaching.} 

\begref{References}

\ref Abt, H.A., 1980, PASP 92, 249 
\ref Abt, H.A., 1981, PASP 93, 207
\ref Abt, H.A., 1984, PASP 96, 563
\ref Abt, H.A., 1985, PASP 97, 1050
\ref Abt, H.A., 1987, PASP 99, 1329
\ref Davoust, E., Schmadel, L., 1987, PASP 99, 700
\ref Davoust, E., Bercegol, H, Callon, M., 1993, {\it Les Cahiers de l'ADEST}, 
special issue, July 1993, p. 44
\ref de Vaucouleurs, G., 1993, ApJ 415, 10 
\ref de Vaucouleurs, G., Bollinger, G., 1979, ApJ 233, 433
\ref Jaschek, C., 1992, Scientometrics, 23, 377
\ref Peterson, C.J., 1987, J. Roy. Astron. Soc. Can. 81, 30
\ref Peterson, C.J., 1988, PASP 100, 106
\ref Trimble, V., 1984, PASP 96, 1007
\ref Trimble, V., 1985, Q. J. R. astr. Soc. 26, 40
\ref Trimble, V., 1993a, Q. J. R. astr. Soc. 34, 235
\ref Trimble, V., 1993b, Q. J. R. astr. Soc. 34, 301
\ref Vin, M.-J., 1995, DESS de gestion de l'information scientifique et 
technique, Universit\'e d'Aix-Marseille III, centre de St J\'er\^ome.

\endref
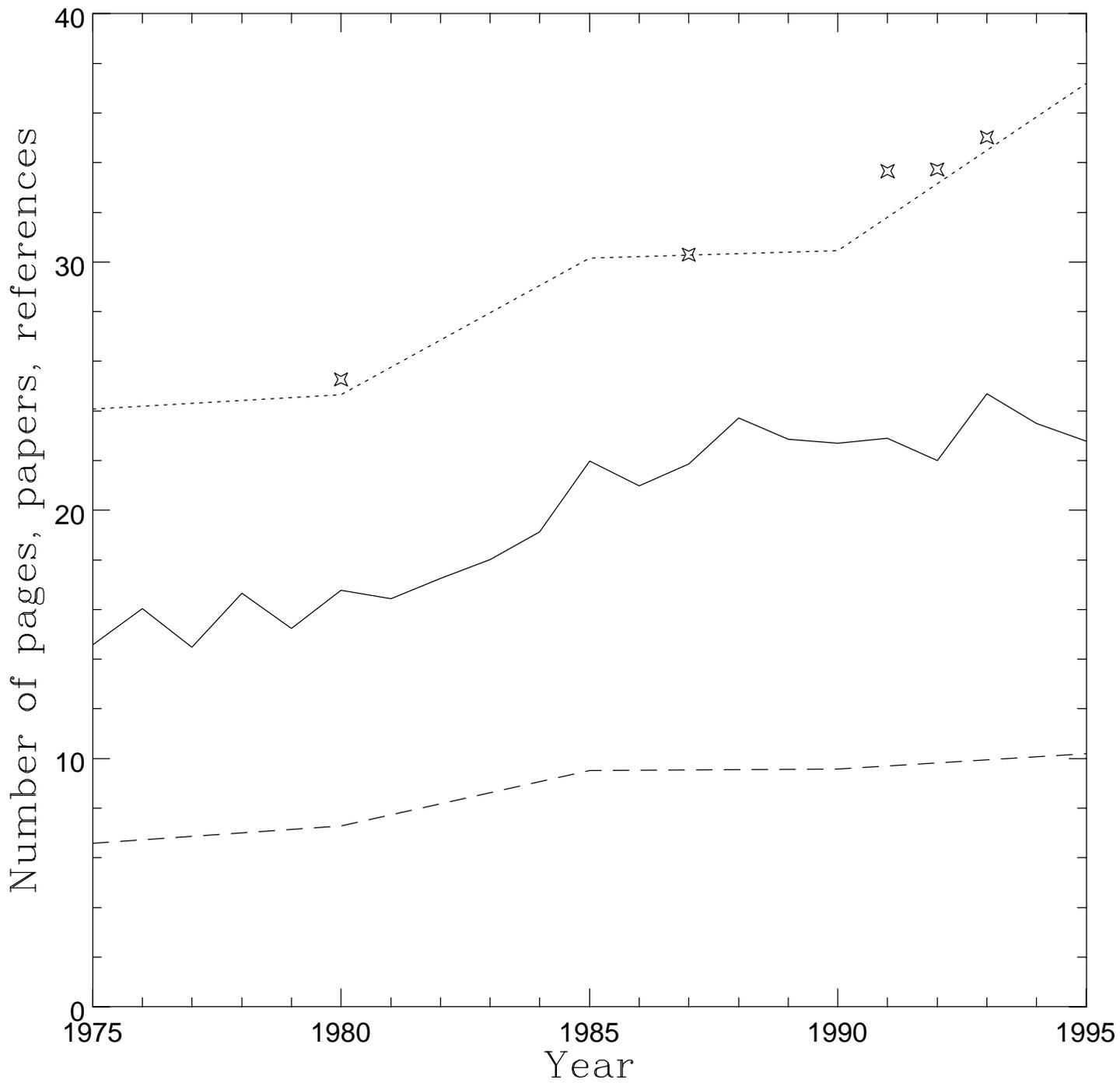
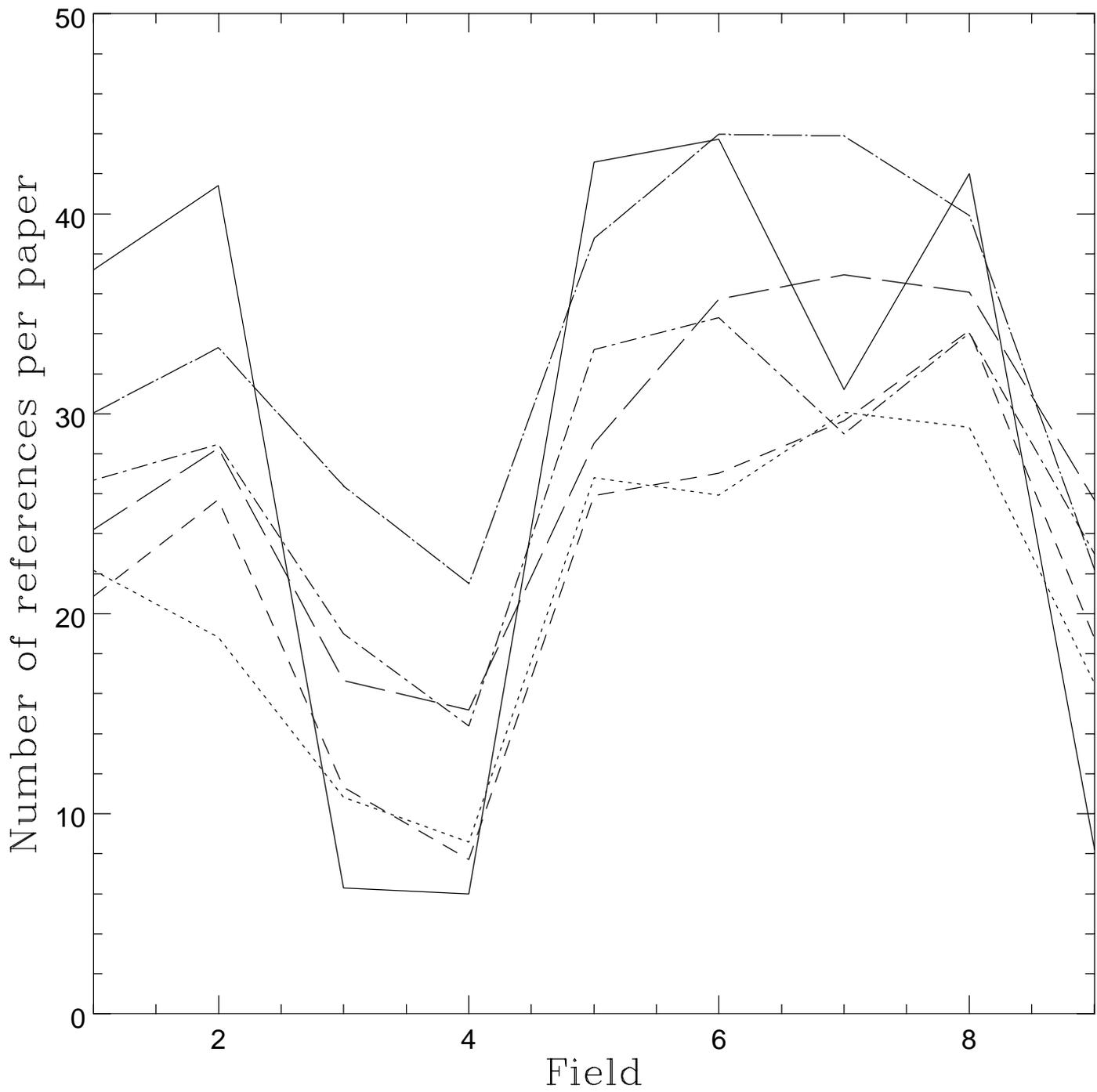
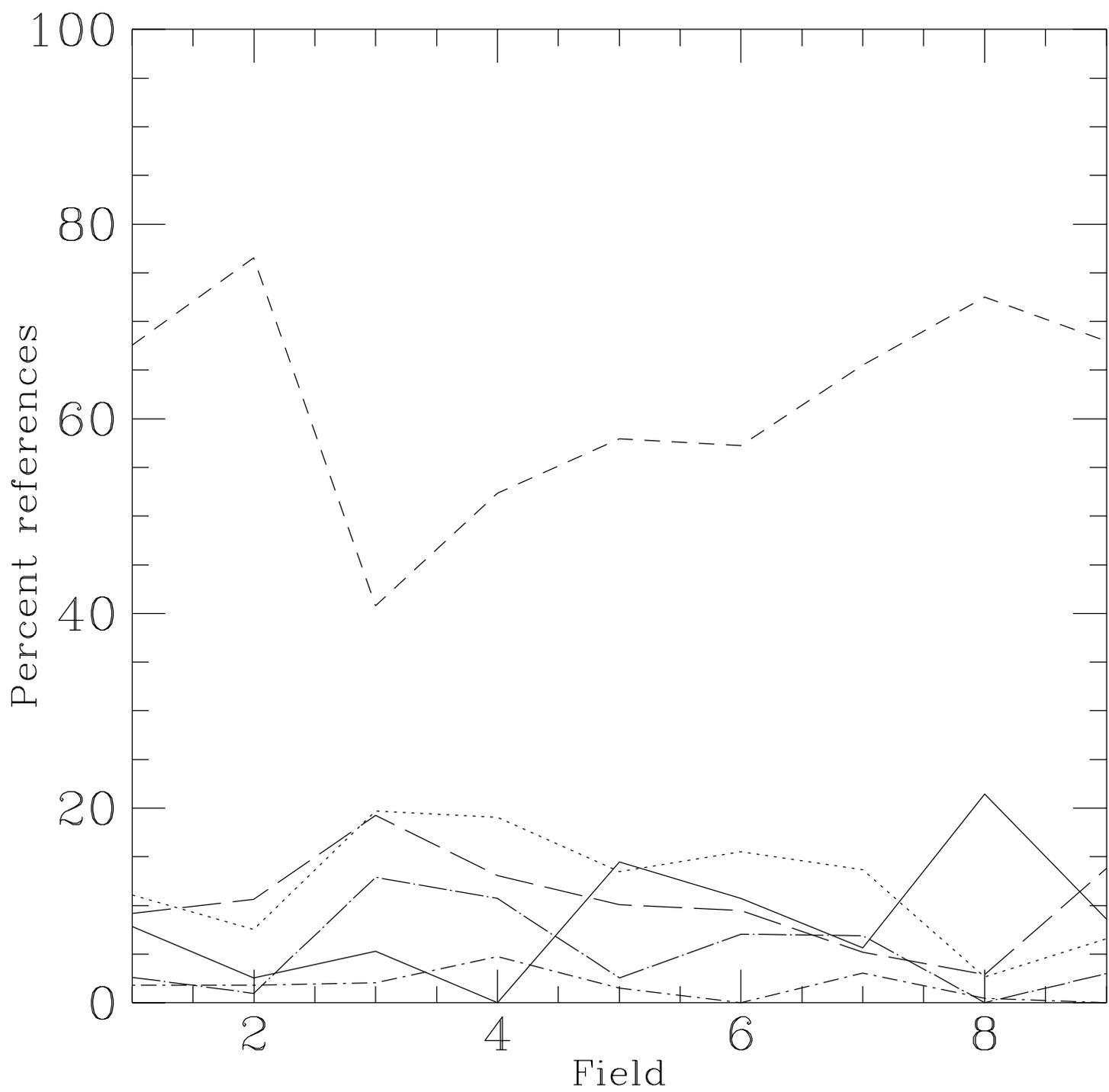
\bye